\documentclass[twoside, a4paper, 10pt]{article}
\usepackage[utf8]{inputenc}
\usepackage[T1]{fontenc}
\usepackage{RR}
\usepackage[numbers]{natbib}
\usepackage{microtype}
\RRtitle{Extensions du langage C pour la programmation hybride CPU/GPU avec StarPU}
\RRetitle{C Language Extensions for Hybrid CPU/GPU Programming with StarPU}
\RRauthor{Ludovic Courtès}
\RCBordeaux
\RRprojet{Runtime}
\RRkeyword{parallel programming, GPU, scheduling, programming language support}
\RRmotcle{programmation parallèle, GPU, ordonnancement, langage de programmation}
\RRNo{8278}

\RRdate{April 2013}

\begin{document}

\RRabstract{

Modern platforms used for high-performance computing (HPC) include
machines with both general-purpose CPUs, and ``accelerators'', often in
the form of graphical processing units (GPUs).  StarPU is a C library
to exploit such platforms.  It provides users with ways to
define \emph{tasks} to be executed on CPUs \emph{or} GPUs, along with the
dependencies among them, and by automatically scheduling them over \emph{all}
the available processing units.  In doing so, it also relieves
programmers from the need to know the underlying architecture details:
it adapts to the available CPUs and GPUs, and automatically transfers
data between main memory and GPUs as needed.

While StarPU's approach is successful at addressing run-time scheduling
issues, being a C library makes for a poor and error-prone programming
interface.  This paper presents an effort started in 2011
to promote some of the concepts exported by the library as C language
constructs, by means of an extension of the GCC compiler suite.  Our
main contribution is the design and implementation of language
extensions that map to StarPU's task programming paradigm.  We argue
that the proposed extensions make it easier to get started with StarPU,
eliminate errors that can occur when using the C library, and
help diagnose possible mistakes.  We conclude on future work.

}

\RRresume{
Les plateformes modernes utilisées en calcul intensif (HPC) incluent des
machines comprenant à la fois des unités de traitement généralistes
(CPU) et des ``accélérateurs'', souvent sous la forme d'unités de
traitement ``graphiques'' (GPU).  StarPU est une bibliothèque C pour
programmer sur ces plateformes.  Elle fournit aux utilisateurs des
moyens de définir des \emph{tâches} pouvant s'exécuter aussi bien sur
CPU que sur GPU, ainsi que les dépendances entre ces tâches, et s'occupe
de les ordonnancer sur \emph{toutes} les unités de traitement
disponibles.  Ce faisant, StarPU abstrait le programmeur des détails
techniques sous-jacents: StarPU s'adapte aux unités de traitement
disponibles et se charge de transférer les données entre elles quand
cela est nécessaire.

StarPU traite efficacement des problèmes d'ordonnacement, mais
l'interface en langage C qu'elle propose est pauvre et facilite les
erreurs de programmation.  Cet article présente des travaux démarrés en
2011 pour promouvoir certains concepts exposés par la bibliothèque
StarPU sous forme d'extensions du langage C, par le biais d'une
extensions de la suite de compilateurs GCC.  Notre principale
contribution est la conception et la mise en œuvre d'extensions du
langage C correspondant au paradigme de programmation par tâches de
StarPU.  Nous montrons que les extensions proposées facilitent la
programmation avec StarPU, éliminent des erreurs de programmation
pouvant intervenir lorsque la bibliothèque C est utilisée et aident le
diagnostique de possibles erreurs.  Nous concluons sur les travaux à
venir.
}

  \makeRR
  \clearpage
  \tableofcontents

\section{Introduction}
\label{sec-1}

Exploiting modern machines that include several CPU cores and several
GPUs is a tough problem.  First, because GPUs offer parallel processing
capabilities different from that of general-purpose CPUs, and are
typically programmed in a specific language such as OpenCL.  Second,
even though GPUs perform better than CPUs for a number of algebraic
computations commonly found in numerical simulation software, CPUs
perform better for some particular computations, and, more importantly,
the number of CPU cores per machine has kept increasing.  Thus, CPUs
\emph{still} have a role to play in improving the performance of HPC
software, as evidenced by Augonnet et al. in
\cite{AugThiNamWac09Europar}.  Consequently, today's challenge is the
exploitation of \emph{all} the available processing units.  Third, HPC
software should be able to perform adequately on any machine with CPUs
and accelerators, without requiring machine-specific adjustments---in
other words, \emph{performance portability} should be achieved.  Finally, the
details of how to program these architectures should be hidden to
programmers, so they can instead focus on higher-level programming
tasks.

StarPU is an effort to address these problems by providing a uniform
run-time support for HPC applications, in the form of a C library
\cite{AugThiNamWac09Europar}.  The library allows programs to define
tasks that may run either on CPUs or GPUs, and to express data
dependencies among these tasks.  This gives StarPU a (partial) view of the
program's directed acyclic graph (DAG) of tasks, which it then schedules on
the available processing units.  StarPU implements scheduling algorithms
from the literature, such as the \emph{heterogeneous earliest finish time}
(HEFT) algorithm \cite{Topcuouglu:2002:PLT:566137.566142}, which allows
it to achieve good performance on hybrid CPU/GPU machines for widespread
linear algebra computations
\cite{AguAugDonLtaNamRomThiTom10SAAHPC,AguAugDonFavLanLtaTomAICCSA11,AguAugDonFavLtaThiTom11IPDPS}.
StarPU takes care of transferring data between main memory and GPUs as
needed, thereby relieving programmers from having to care about these
low-level details.

In the next section, we present StarPU's task-based programming paradigm
for hybrid CPU/GPU targets.  We then describe how we promoted concepts
of this programming paradigm as C language constructs.  We compare to
related work, and conclude on future work.
\section{StarPU's Task Programming Model}
\label{sec-2}

StarPU schedules user-provided \emph{tasks} over the available processing
units.  User tasks may have several \emph{implementations}.  For instance,
a task may have a CPU implementation written in C, an additional CPU
implementation also written in C but using SIMD ISA extensions such as
SSE, and a GPU implementation written in OpenCL.

Conceptually, tasks are functions with scalar arguments
and \emph{buffer} parameters.  ``Buffer'' parameters denote large pieces of data
that may have to be transferred back and forth between main memory and
GPUs, and that may be accessed read-only, write-only, or read-write
by the task.  These access modes, along with the sequence of task
invocations, allows StarPU to determine at run-time the \emph{dependency graph of tasks} \cite{AugThiNamWac09Europar}.

The C programming interface can be used as follows.  First, a
\texttt{starpu\_codelet} structure must be defined.  It describes the task, its
implementations, and its parameters:

\begin{verbatim}
void scale_vector_cpu (void *buffers[], void *args);
void scale_vector_opencl (void *buffers[], void *args);

static struct starpu_codelet scale_vector_codelet =
{
   .cpu_funcs = { scale_vector_cpu, NULL },
   .opencl_funcs = { scale_vector_opencl, NULL },
   .nbuffers = 1,
   .modes = { STARPU_RW },
   .name = "scale_vector"   /* for debugging purposes */
};
\end{verbatim}

The above code defines a task with one CPU and one OpenCL (GPU)
implementation.  This task has one formal parameter, which is a
read-write (or input/output) buffer.  The actual CPU task implementation
is then defined like this:

\begin{verbatim}
void scale_vector_cpu (void *buffers[], void *arg)
{
   /* Unpack the arguments... */
   float *factor = arg;
   starpu_vector_interface_t *vector = buffers[0];
   unsigned n = STARPU_VECTOR_GET_NX (vector);
   float *val = (float *) STARPU_VECTOR_GET_PTR (vector);

   /* scale the vector */
   for (unsigned i = 0; i < n; i++)
     val[i] *= *factor;
}
\end{verbatim}

The code above casts the scalar argument to a float, and the untyped pointer
to the actual vector of \texttt{float} that the task expects\footnote{When several scalar arguments are passed, they have to be
  \emph{unmarshalled} from \texttt{arg}, using the \texttt{starpu\_codelet\_unpack\_args}
  helper.
 }.  The actual
computation follows, accessing the vector directly.  The OpenCL
implementation of this task would be along these lines:

\begin{verbatim}
void vector_scal_opencl (void *buffers[], void *arg)
{
  /* ... */
  err = starpu_opencl_load_kernel (&kernel, &queue, &cl_programs,
                                   "vector_scal_opencl", devid);
  err = clSetKernelArg (kernel, 0, sizeof (val), &val);
  err |= clSetKernelArg (kernel, 1, sizeof (size), &size);
  /* ... */
  err = clEnqueueNDRangeKernel (queue, kernel, 1, NULL, &global,
                                &local, 0, NULL, &event);
  /* ... */
  clFinish (queue);
  /* ... */
}
\end{verbatim}

This is the usual boilerplate that one would need to load an OpenCL
``kernel'' from its OpenCL source file, to prepare its invocation, and to
enqueue it on an OpenCL device (which may be either a GPU or a CPU).
All these constitute a complete StarPU task definition.

The task invocation involves two steps: registering memory buffers that
will be passed to tasks, and actually invoking the task.  The former is
a necessary step: it allows StarPU's memory management component to know
which data buffers are used, and to transfer them as needed.  Then the
\texttt{starpu\_insert\_task} function makes an asynchronous call to the task:

\begin{verbatim}
starpu_data_handle_t vector_handle;
starpu_vector_data_register (&vector_handle, 0, vector,
                             NX, sizeof (vector[0]));

float factor = 3.14;

starpu_insert_task (&scale_vector_codelet,
                    STARPU_VALUE, &factor, sizeof factor,
                    STARPU_RW, vector_handle,
                    0);

/* ... */
starpu_task_wait_for_all ();      /* wait for task completion */
starpu_data_unregister (vector_handle);
\end{verbatim}

As can be seen, the standard C API is a poor way to express the concepts
associated with StarPU's task programming paradigm.  It leads to verbose
code, even encouraging users to duplicate boilerplate, such
as the OpenCL kernel invocation.

In addition to being verbose, it is also error-prone: the arguments
passed to \texttt{starpu\_insert\_task} must be consistent with what
\texttt{scale\_vector\_cpu} and \texttt{scale\_vector\_opencl} expect; failing to do so
will obviously result in undefined behavior, without warnings.
Finally, the API forces users to deal with concepts such as ``data
handles'', which are really StarPU's internal concern.
\section{Blending StarPU Tasks into C}
\label{sec-3}

We believe that StarPU's programming interface calls for language and
compiler support.  The definition of tasks and their implementations, and
the invocation of tasks, ought to be similar to standard function
definitions and invocations.  Memory management ought to impose as
little burden as possible on programmers.

With this in mind, we extend C-family languages, \emph{via} a plug-in for the
GNU Compiler Collection (GCC), with support to directly express these
concepts.  Our extensions define \emph{annotations} that can be added to a
standard C program, turning it into a StarPU program; compiling the
annotated program without StarPU's compiler plug-in still leads a valid
sequential program.  The GCC plug-in is part of the StarPU package since version
1.0.0, released in March 2012\footnote{See \href{http://runtime.bordeaux.inria.fr/StarPU/}{http://runtime.bordeaux.inria.fr/StarPU/} for software
  downloads.  See \href{http://gcc.gnu.org/ml/gcc/2012-03/msg00457.html}{http://gcc.gnu.org/ml/gcc/2012-03/msg00457.html} for
  the original announcement of StarPU 1.0.0.
 }.

This section motivates our choice of GCC as the target platform, and
describes our language extensions and their implementation as well as
the additional benefits they provide.
\subsection{Extending the GNU Compiler Collection}
\label{sec-3-1}

The choice of incorporating support into GCC, as opposed to using a
source-to-source compiler, was motivated by the following reasons.
First, GCC being widely available, implementing our language extensions
as a plug-in means that they would be more readily usable by a number of
users.  Second, GCC contains robust implementations of the C, C++, and
Objective-C languages, all of which can use our extensions; among the
other language front-ends of interest to HPC programmers is Fortran,
though supporting it would have required adjustments to our plug-in that
we did not make.  GCC also supports known language extensions for
parallel programming, such as OpenMP, Cilk Plus, and UPC.  Conversely,
source-to-source compilers often come with incomplete language
front-ends, and often with fewer of them.

Third, implementing this support \emph{within} GCC has allowed us to
work not only on the front-end level, but also at lower levels: our
implementation takes advantage of the \texttt{GENERIC}, \texttt{GIMPLE}, and
\texttt{Tree-SSA} intermediate representations and associated facilities, as
described below.  Tight integration with an optimizing compiler also
enables better code generation, avoiding the traps associated with
implementations of parallel programming directives as compiler
front-ends \cite{pop:inria-00551518}.
\subsection{Defining Tasks and Their Implementations}
\label{sec-3-2}

Our C extensions rely on the two syntactic mechanisms used for C-family
language extensions in GCC: pragmas, and attributes
\cite{stallman13:gcc-manual}.  Attributes are a now widely-adopted GNU C
extension that allows annotations to be associated with individual
elements of the abstract syntax tree (AST).  It uses a syntax compatible
with that of C; as such it may be used in macros, and attribute
arguments may refer to identifiers or operators of the C language.
Conversely, pragmas extend the C pre-processor syntax\footnote{C99 introduced the \texttt{\_Pragma} construct, which extends the syntax
  of C rather than that of the pre-processor, making it amenable to use
  in macros.  Like \texttt{\#pragma}, its sole argument is a string whose syntax
  and semantics are not necessarily related to those of the C language.
 }, and
essentially introduce a new local syntax.  Unknown attributes and
unknown pragmas are simply ignored by the compiler---a property that
makes programs using our extensions valid sequential programs when
compiler support is missing.

We define a \texttt{task} attribute for task declarations.  Tasks are
\texttt{task}-qualified C functions whose return type is \texttt{void}.  The access
mode of its parameters is determined based on their type and qualifiers.
Scalar parameters are passed by value.  Pointer parameters are
considered to be \emph{read-write} when they are not qualified, \emph{read-only}
when they are \texttt{const}-qualified, and \emph{write-only} when they are
qualified with \texttt{\_\_attribute\_\_ ((output))}.

The \texttt{task\_implementation} attribute allows a C function to be
declared as the implementation of a task for a particular target:

\begin{verbatim}
void scale_vector (int size, float vector[size],
                   float factor)
  __attribute__ ((task));

void scale_vector_cpu (int size, float vector[size],
                       float factor)
  __attribute__ ((task_implementation ("cpu", scale_vector)));
\end{verbatim}

The first argument of \texttt{task\_implementation} is a string identifying the
target, one of \texttt{cpu}, \texttt{opencl}, or \texttt{cuda}.  The second argument is the
identifier of the task being implemented.  Each task implementation must
be defined, either in the same compilation unit or in a different one.
The compiler emits helpful error messages when those attributes are used
inappropriately---e.g., when they are used on a non-function, or when
there is a signature mismatch between a task and its implementation.

Tasks themselves must not be defined by the user; instead, their body
is automatically generated by the compiler.  For instance, the
generated body of \texttt{scale\_vector} above is along these lines:

\begin{verbatim}
void
scale_vector (unsigned int size, float *vector, float factor)
{
  starpu_data_handle_t handle;
  int D.10983;
  char * D.10984;
  int err;

  handle = starpu_data_lookup (vector);
  if (handle == 0B) goto <D.10979>; else goto <D.10980>;
  <D.10979>:
  __builtin_puts (&"scale_vector.c:36: error: \
attempt to use unregistered pointer\n"[0]);
  __builtin_abort ();
  <D.10980>:
  err = starpu_insert_task (&scale_vector.codelet, 16, &size,
                            4, 3, handle, 16, &factor, 4, 0);
  if (err != 0) goto <D.10981>; else goto <D.10982>;
  <D.10981>:
  D.10983 = -err;
  D.10984 = strerror (D.10983);
  __builtin_printf (&"scale_vector.c:36: error: failed \
to insert task `scale_vector\': %s\n"[0], D.10984);
  __builtin_abort ();
  <D.10982>:
}
\end{verbatim}

As can be seen above, the generated body of the task itself does
essentially two things: it retrieves the memory handle that the \texttt{vector}
pointer corresponds to, and submits the task for execution with
\texttt{starpu\_insert\_task}.  Error-checking code is inserted as well.  As with
OpenMP, user-defined error handling is not possible.  This limitation is
mostly the result of the design criterion to have code that remains
valid sequential code modulo annotations, coupled with lack of support
for exceptions in C.

Note that task implementations are directly ``human-readable''---an
improvement over code written against StarPU's standard C API.  For each
task implementation, a \emph{wrapper} is generated that takes care of
marshalling/unmarshalling arguments as passed by StarPU's run-time
support:

\begin{verbatim}
void
scale_vector_cpu.task_impl_wrapper (void *buffers[], void *cl_args)
{
  void * D.10999;
  float scalar_arg.30;
  unsigned int scalar_arg.31;
  unsigned int scalar_arg.25;
  float * pointer_arg.26;
  float scalar_arg.27;

  D.10999 = *buffers.28;

  /* STARPU_VECTOR_GET_PTR */
  pointer_arg.26 = MEM[(float * *)D.10999];

  starpu_codelet_unpack_args (cl_args,
                              &scalar_arg.25, &scalar_arg.27);
  scalar_arg.30 = scalar_arg.27;
  scalar_arg.31 = scalar_arg.25;
  scale_vector.cpu_implementation (scalar_arg.31, pointer_arg.26,
                                   scalar_arg.30);
}
\end{verbatim}

As a shorthand, the C extensions support \emph{implicit CPU task implementation declarations}: when the user provides the body of a
\texttt{task}-qualified function, the compiler assumes that the body is that of
the CPU implementation of that task, and does all the necessary
rewriting.

Declaring a task leads to the declaration of the corresponding
\texttt{starpu\_codelet} structure.  Now, how can we determine the compilation
unit in which to emit the \emph{definitions} of the \texttt{starpu\_codelet}
structure, task implementation wrappers, and task body?

The choice we made is that, if one or more task implementations are
\emph{defined} in a compilation unit, then the \texttt{starpu\_codelet} structure
definition, the task implementation wrapper, and the task's body all get
defined in that compilation unit.  Consequently, task implementations
are expected to be defined in the same file.  One exception is CUDA task
implementations: these are normally written in CUDA, not in C, and will
obviously be defined in a file of their own; their declaration as a
\texttt{task\_implementation} must be visible in the compilation unit where the
other task implementations are defined.
\subsection{Invoking Tasks}
\label{sec-3-3}

Tasks invocations have the same syntax as regular C function calls.  For
instance, the \texttt{scale\_vector} task defined above may be invoked like
this:

\begin{verbatim}
float vector[NX];
/* ... */
scale_vector (NX, vector, 3.14);
\end{verbatim}

An important difference with standard C is that task invocations are
\emph{asynchronous}: the invocation statement just adds the task call to the
scheduler's queue, which will pick it up eventually, and possibly
execute it on a different thread.  Programs can wait for the completion
of all pending task invocations using \texttt{\#pragma starpu wait}.

As described before, the body of the \texttt{scale\_vector} task does the actual
queueing, using the \texttt{starpu\_insert\_task} function.
\subsection{Expressing Memory Management}
\label{sec-3-4}

As explained earlier, memory buffers passed as task arguments must be
registered to StarPU, so that they can be transferred back and forth
between main memory and GPUs when appropriate.  The C language
extensions provide pragmas that directly reify the corresponding C
functions: \texttt{\#pragma starpu register} translates to a call to
\texttt{starpu\_vector\_data\_register}, \texttt{\#pragma starpu unregister} translates to
a call to \texttt{starpu\_data\_unregister}, etc.

The added convenience over the C interface is improved error checking,
and conciseness.  For instance, when registering a variable that has
array type with a known size, users need not specify the array size:

\begin{verbatim}
int
foo (void)
{
  static float global[123];

#pragma starpu register global
  /* ... */
}
\end{verbatim}

Other programming errors are avoided.  For instance, attempts to
register an automatic variable lead to a compile-time warning noting
that the storage of these variables may be reclaimed before tasks that
use it have completed.

StarPU's GCC plug-in also introduces two extensions for lexically-scoped
dynamic memory allocation and registration.  Again, the goal is to allow
for concise code, and to avoid common programming errors related to
manual memory management.  Two new attributes are defined: the
\texttt{heap\_allocated} attribute marks an array-typed variable as having
storage allocated on the heap, and the \texttt{registered} attribute marks an
array-typed variable has being registered for use with StarPU tasks.

Both attributes have block scope: they take effect at the variable
definition point, and are undone when the variable's scope is
left---just like C++ automatic variables.  They are typically used
together, as follows:

\begin{verbatim}
int
func (void)
{
  int matrix[123][234][77]
    __attribute__ ((registered, heap_allocated));

  /* ... */

  some_task (matrix);
#pragma starpu wait

  /* Make sure MATRIX is available in main memory.  */
#pragma starpu acquire matrix

  dump_matrix (matrix, 123, 234, 77);

  /* MATRIX is unregistered and deallocated here.  */
}
\end{verbatim}

Behind the scenes, the \texttt{heap\_allocated} attribute leads to the
generation of a \texttt{starpu\_malloc} call and corresponding \texttt{starpu\_free}
call.  The \texttt{starpu\_malloc} function works like \texttt{malloc}, but it also
tries to pin the allocated memory in CUDA or OpenCL, so that data
transfers from this buffer can be asynchronous, thereby permitting
data transfer and computation overlapping.  GCC's \texttt{GENERIC} intermediate
representation supports \emph{cleanup handlers}, which is what is used here
to guarantee that \texttt{starpu\_free} is called when the variable's scope is
left.

Finally, StarPU's GCC plug-in leverages the static analysis
infrastructure available in GCC to warn against possible omissions of a
\texttt{registered} attribute or \texttt{register} pragma.  Here's an example:

\begin{verbatim}
extern void my_task (size_t a, double *x, size_t b, double *y)
  __attribute__ ((task));

void
one_unregistered_pointer (void)
{
  double *p, *q;

  p = malloc (12 * sizeof *p);
  q = malloc (23 * sizeof *q);

#pragma starpu register p 12
  my_task (12, p, 23, q); /* <- warning here */
}
\end{verbatim}

Compiling this code yields a warning on the \texttt{my\_task} call:

\begin{verbatim}
example.c:10:11: warning: variable 'q' may be used unregistered
\end{verbatim}

This is achieved by working on the SSA form of the code, looking for
calls to \texttt{starpu\_vector\_data\_register}, and then checking the arguments
of any such call.  If the argument of a dominating
\texttt{starpu\_vector\_data\_register} call is found to alias the memory region
pointed to by the argument of interest, then no warning is raised;
otherwise, a warning is emitted, with the name of the variable.
\subsection{OpenCL support}
\label{sec-3-5}

Our GCC plug-in helps with the integration of OpenCL kernels in two
ways.  First, the GCC plug-in provides a helper for the generation of task
implementations that simply launch an OpenCL kernel.  Writing the C code
that loads an OpenCL source file, builds a particular kernel from that
file, and submits it for execution on an OpenCL device is cumbersome.
Using the standard OpenCL API along with StarPU's C library helpers, the
body of an OpenCL task implementation looks like this
\cite{runtime12:starpu-handbook}:

\begin{small}

\begin{verbatim}
static void vector_scal_opencl (unsigned size, float vector[size],
                                float factor)
  __attribute__ ((task_implementation ("opencl", vector_scal)));

static void
vector_scal_opencl (unsigned size, float vector[size], float factor)
{
  int id, devid, err;
  cl_kernel kernel;
  cl_command_queue queue;
  cl_event event;

  /* VECTOR is GPU memory pointer, not a main memory pointer.  */
  cl_mem val = (cl_mem) vector;

  id = starpu_worker_get_id ();
  devid = starpu_worker_get_devid (id);

  /* Prepare to invoke the kernel.  In the future, this will be largely
     automated.  */
  err = starpu_opencl_load_kernel (&kernel, &queue, &cl_programs,
                                   "vector_mult_opencl", devid);
  if (err != CL_SUCCESS)
    STARPU_OPENCL_REPORT_ERROR (err);

  err = clSetKernelArg (kernel, 0, sizeof (val), &val);
  err |= clSetKernelArg (kernel, 1, sizeof (size), &size);
  err |= clSetKernelArg (kernel, 2, sizeof (factor), &factor);
  if (err)
    STARPU_OPENCL_REPORT_ERROR (err);

  size_t global = 1, local = 1;
  err = clEnqueueNDRangeKernel (queue, kernel, 1, NULL, &global,
                                &local, 0, NULL, &event);
  if (err != CL_SUCCESS)
    STARPU_OPENCL_REPORT_ERROR (err);

  clFinish (queue);
  starpu_opencl_collect_stats (event);
  clReleaseEvent (event);

  /* Done with KERNEL.  */
  starpu_opencl_release_kernel (kernel);
}
\end{verbatim}
\end{small}

In addition, the OpenCL source code itself must be loaded elsewhere, for
instance from \texttt{main}:

\begin{verbatim}
starpu_opencl_load_opencl_from_file ("vector_scal_opencl_kernel.cl",
                                     &cl_programs, "");
\end{verbatim}

All this is a cumbersome, error-prone, and repetitive task.
Additionally, this approach requires that the OpenCL source file of
interest be available in the current directory \emph{at run time}---another
inconvenience.

Based on this experience, we added an \texttt{opencl} pragma to automate this
task.  It is used like this:

\begin{verbatim}
static void my_task (int x, float a[x])
  __attribute__ ((task));

static void my_task_opencl (int x, float a[x])
  __attribute__ ((task_implementation ("opencl", my_task)));

#pragma starpu opencl my_task_opencl "my-kernel.cl" "kern" 8
\end{verbatim}

As can be seen, the task and its implementation must still be declared.
At the point where the \texttt{opencl} pragma is used, the body of
\texttt{my\_task\_opencl} is generated.  The generated code is similar to what we
shown above: it enqueues the OpenCL function \texttt{kern}, defined in
\texttt{my-kernel.cl}, for execution on the OpenCL device chosen by StarPU's
scheduler, and with a group size equal to \texttt{8}.  The source code from
\texttt{my-kernel.cl} is actually read \emph{at compile time}, and stored in a
generated global variable; the generated code then simply loads the
OpenCL program from that string.  This eliminates the need to have
\texttt{my-kernel.cl} available at run time.

We believe this simple facility provides a practical benefit for writers
of heterogeneous applications in C and OpenCL with StarPU.  Another
interesting approach would be the automatic generation of OpenCL kernels
from C, as performed by HMPP \cite{dolbeau07:hmpp} and OpenACC
\cite{openacc1.0:2012} (see the \hyperref[sec-5]{``Future Work'' section} for a discussion.)
However, automatic kernel generation may only be applicable to a
restricted set of input C functions \cite{kravets2009:graphite-opencl},
and it may be difficult to generate kernels as efficient as hand-written
ones.  Consequently, we believe that using hand-written OpenCL kernels
remains relevant.

Second, our GCC plug-in raises a warning when a task with an OpenCL
implementation uses parameter types that do not exist in OpenCL, or that
have a different definition, as in this example:

\begin{verbatim}
static void my_task (size_t size, int x[size])
  __attribute__ ((task));

static void my_task_opencl (size_t size, int x[size])
  __attribute__ ((task_implementation ("opencl", my_task)));
\end{verbatim}

Since \texttt{size\_t} does not exist in OpenCL, the actual kernel necessarily
uses a different type, which may be incompatible.  Thus, the following
warning is emitted:

\begin{verbatim}
warning: 'size_t' does not correspond to a known OpenCL type
\end{verbatim}

Likewise, other scalar types exist both in OpenCL and standard C, but
with a different definition.  For instance, the OpenCL specification
defines \texttt{cl\_long} to be a 64-bit signed integer type.  Consider this
example:

\begin{verbatim}
static void my_long_task (long size, int x[size])
  __attribute__ ((task));

static void my_long_task_opencl (long size, int x[size])
  __attribute__ ((task_implementation ("opencl", my_long_task)));
\end{verbatim}

Compiling this on a 32-bit platform, where \texttt{long} in C is a 32-bit type,
yields the following warning:

\begin{verbatim}
warning: C type 'long int' differs from the same-named OpenCL type
\end{verbatim}

The same goes for types differing in signedness, such as \texttt{char}.  This
helps avoid simple but possibly hard-to-track programming errors when
using OpenCL kernels from C.
\section{Related Work}
\label{sec-4}

This section comments on related work in the area of programming
language extensions, for C and related languages, for heterogeneous
programming.
\subsection{Jade}
\label{sec-4-1}

While the idea of programming heterogeneous systems has become prevalent
over the last few years as GPU became popular, earlier work had been
done on this topic.  The Jade project
\cite{Rinard:1992:HPP:147877.148003,Rinard:1998:DIE:291889.291893}
addressed the problem of heterogeneous parallel programming and C
language extensions starting from 1992.

The targeted hardware back then was not GPUs, but instead networks of
heterogeneous workstations, message-passing machines, and SMP machines.
Users of Jade's C extensions must start from a sequential C program,
split it into tasks suitably, specify the access modes of buffer
parameters (called \emph{shared objects}), and describe how data are to be
decomposed in atomic units actually accessed by the program (using \emph{part objects}).

Unlike our C extensions, Jade's include new keywords and new syntax,
which improves expressiveness at the expense of making Jade programs not
compilable by standard C compilers.  Jade's run-time support then
distributes tasks across machines, and takes care of any necessary data
transfers \cite{Rinard:1992:HPP:147877.148003}.
\subsection{HMPP}
\label{sec-4-2}

HMPP (for \emph{Hybrid Multicore Parallel Programming}) has been developed at
Inria and then CAPS Entreprise since 2007 \cite{dolbeau07:hmpp}.  HMPP
supports a task-based programming paradigm for heterogeneous machines,
similar to that of StarPU.  At its core is a set of programming
directives that extend the C and FORTRAN languages, similar in spirit to
ours.

The \texttt{codelet} pragma allows programmers to mark a function as being a
candidate to run on a GPU---a \emph{codelet}---and additional clauses can be
used to specify whether pointer or array arguments are used as input,
output, or both; the \texttt{target} clause specify whether to use CUDA or some
other GPU-supporting environment as the back-end.  Codelet call sites
must be annotated with the \texttt{callsite} pragma.  It allows users to
provide information such as the size of arrays passed as arguments to
the codelet, and whether the codelet invocation is synchronous or
asynchronous.  In addition to annotating functions, HMPP supports the
annotation of \emph{code blocks} directly, \emph{via} the \texttt{region} directive.

One of the main advantages of HMPP is that it generates target GPU code
directly from annotated C or FORTRAN codelets.  This is a relief for
application programmers who no longer need to learn and integrate
different languages.

HMPP provides \texttt{allocate}, \texttt{release}, and other directives for explicit
data transfers between main memory and the GPUs.  Unfortunately, this
removes flexibility to its run-time support, and hinders performance
portability, as has been shown by work on StarPU
\cite{AugThiNamWac09Europar}.  As with OpenACC, it also assumes that a
single GPU is in use.
\subsection{OmpSs}
\label{sec-4-3}

More recently, OmpSs has been developed to address heterogeneous
programming on hybrid CPU/GPU machines as well as clusters thereof
\cite{bueno:2012:ompssgpu}.  As for Jade and StarPU, this work includes
both run-time support for dynamic scheduling, and C language extensions.
OmpSs extensions are based on the \texttt{pragma} mechanism, which allows
OmpSs-annotated programs to remain valid sequential programs, as with
StarPU's C extensions.

The proposed extensions are similar in spirit to StarPU's.  Functions
may be annotated with a \texttt{task} directive with \texttt{input} and \texttt{output}
clauses to specify the task's arguments access modes; along
with the \texttt{concurrent} clause, it allows OmpSs to determine the data
dependencies among task invocations, like StarPU does by default.  Calls
to \texttt{task}-annotated functions are asynchronous.  In addition, the \texttt{task}
pragma may be used to annotate directly a call to a standard C function.

Additionally, OmpSs provides a \texttt{target} pragma for \texttt{task}-annotated
functions, that specifies where the task it to run, and how it is
implemented (for instance, \texttt{cuda}).  Unlike StarPU, it appears that
tasks may have only one target, introduced with the \texttt{device} keyword
\cite{bueno:2012:ompssgpu}.  It is up to the programmer to specify which
of the input and output arguments are to copied to and from the device,
\emph{via} additional \texttt{copy\_in}, \texttt{copy\_out}, and \texttt{copy\_inout} clauses.

OmpSs is implemented using the Mercurium source-to-source compiler,
which supports C and C++ as source languages.  Users must run it before
calling the actual C or C++ compiler.
\subsection{OpenACC}
\label{sec-4-4}

OpenACC is a set of C and Fortran extensions, or \emph{programming directives}, designed to simplify off-loading of tasks to accelerators.
Version 1.0 of the specification was released in November 2011
\cite{openacc1.0:2012}.  It defines a set of functions and compiler
\emph{pragmas} to specify parts of a program whose computation may be
offloaded to GPUs, to transfer data between main memory and the GPUs,
and to synchronize with the execution of those parts---the \emph{computational kernels}.

The \texttt{\#pragma acc kernels} directive is used to identify loop nests that
may run on a GPU; it instructs the compiler to generate code for that
architecture.  For better performance, programmers are required to
explicitly state where data transfers between main memory and the GPU
may occur, using one of the \texttt{copy} clauses.

OpenACC's specification suffers from the same shortcomings as C++ AMP.
First, the ``Scope'' section in version 1.0 specifies that the
specification only addresses machines with one accelerator---something
that does not match reality.  Second, it is very much
offloading-oriented: programmers are expected to manually schedule
computational kernels and associated data transfers on their GPU of
choice---a departure from StarPU's performance portability goal.
\subsection{Unified Parallel C}
\label{sec-4-5}

Unified Parallel C (UPC) is an extension of the C language to support
cluster programming, using the \emph{partitioned global address space} (PGAS)
paradigm \cite{upc1.2-2005}.  UPC was recently extended to support GPU
programming \cite{Chen:2010:UPC:1964536.1964547}.  The extended UPC
compiler is able to translate \texttt{upc\_forall} loops to CUDA kernels.

However, UPC obliges programmers to statically define the distribution
of arrays on the available nodes, which then influences work sharing
\emph{via} the \texttt{upc\_forall} affinity parameter.  Consequently, UPC sacrifices
performance portability to a large extent.
\subsection{XcalableMP}
\label{sec-4-6}

Like UPC, XcalableMP (or XMP) is a PGAS extension for C and Fortran for
programming distributed shared memory systems \cite{lee10:xcalablemp},
recently extended for clusters that include GPUs
\cite{lee12:xcalablemp-gpu}.  XMP provides directives for OpenMP-style
work sharing, such as \texttt{loop} and \texttt{reduction}, along with UPC-style
affinity clauses to specify which node executes each iteration.  Similar
to UPC's \texttt{shared} qualifier \cite{upc1.2-2005}, XMP's \texttt{template},
\texttt{distribute}, and \texttt{align} pragmas allow programmers to map arrays to
cluster nodes.  Code blocks can be turned into OpenMP-style tasks using
the \texttt{task} pragma.

XMP-ACC, the XcalableMP extension for GPU programming, supports an
offloading programming paradigm, similar in spirit to that of OpenACC
\cite{lee12:xcalablemp-gpu}.  For instance, programmers must explicitly
state which objects must be allocated on the GPU, using the \texttt{replicate}
pragma, and when they are to be transferred, with the \texttt{replicate\_sync}
pragma, which hampers performance portability, as already noted.  The
\texttt{loop} construct is extended with an \texttt{acc} clause, which explicitly
instructs the compiler and run-time support to execute the loop on a
GPU.  The compiler automatically generates CUDA code for the loop.
\subsection{C++ AMP}
\label{sec-4-7}

Microsoft Corporation designed and implemented C++ AMP, for \emph{Accelerated Massive Parallelism}, with version 1.0 of the specifications released in
August 2012 \cite{amp1.0:2012}.  AMP is a C++ language extension and
associated library whose purpose is to allow programmers to express
parallelism in data-parallel algorithms in a way that allows them to be
offloaded to accelerators such as GPUs.

Programmers write data-parallel functions in C++; functions that
programmers may want to run on GPUs must carry the \texttt{restrict(amp)}
annotation.  The compiler statically checks that \texttt{restrict}-annotated
functions use only language features supported by GPUs, and generates
target code for them.  In addition, AMP provides mechanisms to describe
arrays and tiles, allowing for adequate data and task partitioning.

Unlike StarPU, but like OpenACC, the AMP run-time presumes that only one
accelerator is going to be used for offloading.  The specification
describes heuristics followed by Microsoft's implementation to choose a
``good'' default accelerator \cite{amp1.0:2012}.  Users are otherwise
invited to use the \texttt{accelerator} class to explicitly choose an
accelerator in their programs---a misfeature that hampers performance
portability \cite{AugThiNamWac09Europar}.  This model appears to be a
serious shortcoming for today's machines, which typically contain many CPU
cores along with several accelerators.
\subsection{XKaapi}
\label{sec-4-8}

XKaapi is another run-time support library for task scheduling over
heterogeneous multi-CPU and multi-GPU machines developed at Inria
\cite{ferreiralima:hal-00735470}.  It has the same goals as StarPU, but
addresses them differently: run-time task scheduling is based on \emph{work stealing}, and tasks are launched using the \texttt{spawn} operator reminiscent
of Cilk.  XKaapi supports recursive task invocations, unlike
StarPU.

XKaapi's main API is in C++.  The \texttt{Task::Signature} template allows
programmers to declare tasks and the access modes of their
parameters---similar in spirit to our \texttt{task} attribute--- and the
\texttt{TaskBodyCPU} and \texttt{TaskBodyGPU} templates are used to define task
implementations---similar to our \texttt{task\_implementation} attribute.  Tasks
are invoked using the \texttt{Spawn} function.  The advantage of C++ is that
these features can be added just by using standard mechanisms, without
having to modify the compiler.  The downside is that it remains
noticeably more verbose than what we achieved within GCC, and prevents
good compile-time error reporting and domain-specific static analysis in
the style of what we implemented.

XKaapi also comes with a prototype programming interface that extends
the C and C++ languages with pragmas, developed in 2011, and packaged as
a source-to-source pre-compiler called \emph{KaCC} and based on Rose
\cite{lementec11:xkaapi-api}.  Again, the usual \texttt{task} pragmas allow
functions to be turned into tasks, with the specified access mode of
their parameters.  An additional \texttt{reduction} access mode is available,
which allows several instances of a task to contribute in parallel to
the same result.  Task invocations are written like normal C function
calls, but they must be embedded in a \texttt{parallel}-annotated block to
actually execute concurrently.  The prototype described in
\cite{lementec11:xkaapi-api} does not address heterogeneous programming,
though, so it does not support multiple task implementations nor memory
management.
\section{Future Work}
\label{sec-5}

StarPU's C programming interface offers many features that are not
currently covered by our C language extensions.  This section describes
additional features that could be supported by the C extensions, as well
as new features that could be usefully added by the compiler.
\subsection{Array Partitioning}
\label{sec-5-1}

StarPU's C API defines \emph{filters}, a mechanism that allows data items to
be partitioned into smaller pieces.  For instance, two-dimensional
matrices registered with \texttt{starpu\_matrix\_data\_register} may be
partitioned into sub-matrices \cite{runtime12:starpu-handbook}.  This
allows programmers to invoke tasks and pass them just a subset of the
initial matrix, while allowing tasks that access different subsets of
the matrix to execute in parallel---a common idiom in linear algebra
algorithms such as Cholesky factorizations.

Once a matrix is partitioned, sub-matrices are accessible with
\texttt{starpu\_data\_get\_sub\_data}.  Within tasks, programmers must make sure to
honor the offset, stride, and similar parameters that map logical
indexes within the matrix to their actual location in memory.

Our C extensions do not provide syntactic support for these operations.
This can usually be worked around by structuring data differently, or
registering various parts of the data separately, but the cost of doing
this is to write code that is less natural and more distant from the
algorithm.

One of the difficulties is that in C, arrays are all assumed to be
stored in contiguous memory regions, regardless of their dimensions,
which is no longer the case with StarPU's partitioned matrices.  The
compiler would consequently need to rewrite accesses to such an array in
a way that honors the aforementioned mapping of indexes to memory
locations.  That does not solve all the problems, though, because a
reference to a partitioned array could still be passed to code that is
unaware of the partitioning, and would thus blindly access it as if it
were stored in a contiguous memory region.  Introducing disjoint types for
partitioned arrays may help avoid this.

Another issue is that C does not have any syntax to designate a
sub-array.  Cilk Plus, which is an extension of C and is implemented
within GCC, defines an \emph{array notation} that adds syntax and general
compiler support for this, and would probably be a good starting point
\cite{intel11:cilkplus}.  Jade's \emph{part objects}
\cite{Rinard:1998:DIE:291889.291893} and C++ AMP's \emph{array views}
\cite{amp1.0:2012} also appear like good sources of inspiration.
\subsection{OpenCL Kernel Generation}
\label{sec-5-2}

Our C extensions noticeably lower the barrier of entry to StarPU.
However, it does not help much in the way of writing hybrid CPU/GPU
programs, in that users are still required to write computational
kernels in distinct languages such as OpenCL or CUDA.

One step in that direction would be to automatically generate OpenCL
kernels from CPU task implementations that contain parallelizable loops,
in a way similar to what HMPP and OpenACC achieve
\cite{dolbeau07:hmpp,openacc1.0:2012}.  Experiments have been conducted
in that area, notably within GCC \cite{kravets2009:graphite-opencl},
showing that simple loops could be usefully converted.  As discussed
before, that is only applicable to a restricted set of C functions, and
can lead to kernels less efficient than hand-written ones.  Yet, it
could be a good starting point for people porting an existing
application to a hybrid CPU/GPU architecture.
\subsection{Compiler-Provided Scheduling Hints}
\label{sec-5-3}

Being initially designed as a run-time support library, StarPU discovers
the DAG of tasks at run time, as the program submits them.  However,
when \texttt{task} annotations are used, the compiler is in a good position to
see the DAG of tasks, or a subset thereof.

This could be exploited in different ways.  For instance, task
priorities could be specified by the user.  In some cases, the compiler
may even have enough information to compute the \emph{upward rank} and
\emph{downward rank} of each task, which could then be fed to StarPU's
HEFT-based task scheduler \cite{Topcuouglu:2002:PLT:566137.566142}.
\subsection{Nested Tasks}
\label{sec-5-4}

With StarPU's C API, programs may register \emph{callbacks}, which allows
them to be notified on the completion of the task
\cite{runtime12:starpu-handbook}.  This is often used as a way to
synchronize a task or set of tasks that are waiting on another task's
result.  However, this mechanism introduces \emph{inversion of control}
(IoC), and essentially forces programmers to write in
continuation-passing style.

It would be more natural if CPU task implementations could instead
invoke tasks and wait for their completion.  The compiler would split
the calling task's body at the point where \texttt{wait} is encountered, such
that its continuation is passed as a callback.
\subsection{StarPU-MPI Support}
\label{sec-5-5}

Given that StarPU knows about the DAG of tasks as well as about data
items that are transferred from task to task, adding support for
clusters of machines seemed like a natural extension.  The StarPU-MPI
library does this, by augmenting the API to support MPI-specific idioms
\cite{augonnet:hal-00725477}.

In a nutshell, StarPU-MPI builds upon an MPI implementation.  Users run
one instance of the StarPU-MPI program on each node of the cluster.  The
\texttt{starpu\_mpi\_initialize} function allows the calling process to know its
MPI \emph{rank}.  In addition to StarPU's normal data registration process,
data items that may be passed as arguments to the tasks have a \emph{home node}, which specifies the MPI node holding its initial value, and to
which the item will eventually return.  Upon registration, each process
registers all the data items of interest with
\texttt{starpu\_vector\_data\_register} and similar; when registering a data item
for whose home node is not the calling process, \texttt{NULL} is passed as the
address of the buffer in the \texttt{starpu\_vector\_data\_register} call.
Finally, the \texttt{starpu\_data\_set\_rank} must be used to specify the ``home
node'' of each data item.

With this additional information, the scheduler in each StarPU-MPI
instance is able to decide where to execute each task.  The default
scheduling algorithm chooses the MPI node that runs the task in a way
that minimizes data transfers.

It would be tempting to extend our GCC plug-in to support StarPU-MPI
programming.  Ideally, adding a \texttt{-fstarpu-mpi} compilation flag would
turn the annotated C program into a StarPU-MPI program that may run on a
cluster.  In practice, additional annotations would be needed to
represent MPI-specific information such as the home node of data items,
and the MPI rank.  Jade \cite{Rinard:1998:DIE:291889.291893}, XcalableMP
\cite{lee10:xcalablemp} and other \emph{partitioned global address space}
languages such as UPC \cite{upc1.2-2005} provide hints as to how data
distribution can be expressed.
\section{Conclusion}
\label{sec-6}

StarPU's run-time support has shown how performance portability on
hybrid CPU/GPU machines can be achieved, while abstracting programmers
from the details of the underlying platform, and relieving them from the
need to worry about scheduling of their program's tasks.  By raising the
level of abstraction to match that StarPU's programming paradigm through
annotations, our C language extensions lower the barrier of entry to
StarPU programming.  Programs carrying those annotations remain valid
sequential C programs when the compiler plug-in is unused, in the spirit
of OpenMP.  We have also shown that the extra knowledge made available
to the compiler allows it to better diagnose possible programming
errors.

Future work includes further lowering the barrier of entry to hybrid
CPU/GPU programming, and leveraging compile-time information to guide
the run-time support.

\bibliographystyle{plain}
\bibliography{starpu-gcc}

\begin{thebibliography}{10}

\bibitem{AguAugDonFavLanLtaTomAICCSA11}
Emmanuel Agullo, C{\'e}dric Augonnet, Jack Dongarra, Mathieu Faverge, Julien
  Langou, Hatem Ltaief, and Stanimire Tomov.
\newblock {LU} factorization for accelerator-based systems.
\newblock In {\em 9th ACS/IEEE International Conference on Computer Systems and
  Applications (AICCSA 11)}, June 2011.

\bibitem{AguAugDonFavLtaThiTom11IPDPS}
Emmanuel Agullo, C{\'{e}}dric Augonnet, Jack Dongarra, Mathieu Faverge, Hatem
  Ltaief, Samuel Thibault, and Stanimire Tomov.
\newblock {QR Factorization on a Multicore Node Enhanced with Multiple GPU
  Accelerators}.
\newblock In {\em {25th IEEE International Parallel \& Distributed Processing
  Symposium (IEEE IPDPS 2011)}}, May 2011.

\bibitem{AguAugDonLtaNamRomThiTom10SAAHPC}
{E}mmanuel {A}gullo, {C}{\'e}dric {A}ugonnet, {J}ack {D}ongarra, {H}atem
  {L}taief, {R}aymond {N}amyst, {J}ean {R}oman, {S}amuel {T}hibault, and
  {S}tanimire {T}omov.
\newblock {D}ynamically scheduled {C}holesky factorization on multicore
  architectures with {GPU} accelerators.
\newblock In {\em {S}ymposium on {A}pplication {A}ccelerators in {H}igh
  {P}erformance {C}omputing ({SAAHPC})}, {K}noxville, USA, July 2010.

\bibitem{augonnet:hal-00725477}
C{\'e}dric Augonnet, Olivier Aumage, Nathalie Furmento, Raymond Namyst, and
  Samuel Thibault.
\newblock {StarPU-MPI}: Task programming over clusters of machines enhanced
  with accelerators.
\newblock In Siegfried~Benkner Jesper Larsson~Tr{\"a}ff and Jack Dongarra,
  editors, {\em {The 19th European MPI Users' Group Meeting (EuroMPI 2012)}},
  volume 7490 of {\em LNCS}, Vienna, Austria, 2012. Springer.

\bibitem{AugThiNamWac09Europar}
C{\'e}dric Augonnet, Samuel Thibault, Raymond Namyst, and Pierre-Andr{\'e}
  Wacrenier.
\newblock {StarPU: A Unified Platform for Task Scheduling on Heterogeneous
  Multicore Architectures}.
\newblock In {\em Proceedings of the 15th International Euro-Par Conference},
  2009.

\bibitem{bueno:2012:ompssgpu}
Javier Bueno, Judit Planas, Alejandro Duran, Rosa~M. Badia, Xavier Martorell,
  Eduard Ayguade, and Jes'us Labarta.
\newblock Productive programming of {GPU} clusters with {OmpSs}.
\newblock In {\em Parallel Distributed Processing Symposium (IPDPS), 2012 IEEE
  26th International}, pages 557--568, May 2012.

\bibitem{Chen:2010:UPC:1964536.1964547}
Li~Chen, Lei Liu, Shenglin Tang, Lei Huang, Zheng Jing, Shixiong Xu, Dingfei
  Zhang, and Baojiang Shou.
\newblock {Unified Parallel C} for {GPU} clusters: language extensions and
  compiler implementation.
\newblock In {\em Proceedings of the 23rd international conference on Languages
  and compilers for parallel computing}, LCPC'10, pages 151--165, Berlin,
  Heidelberg, 2011. Springer-Verlag.

\bibitem{dolbeau07:hmpp}
Romain Dolbeau, Stéphane Bihan, and François Bodin.
\newblock {HMPP™}: A hybrid multi-core parallel programming environment.
\newblock In {\em Proceedings of the Workshop on General Purpose Processing on
  Graphics Processing Units ({GPGPU 2007})}, Boston, Massachussets, {USA},
  October 2007.

\bibitem{ferreiralima:hal-00735470}
Joao~Vicente Ferreira~Lima, Thierry Gautier, Nicolas Maillard, and Vincent
  Danjean.
\newblock {Exploiting Concurrent GPU Operations for Efficient Work Stealing on
  Multi-GPUs}.
\newblock In {\em {24rd International Symposium on Computer Architecture and
  High Performance Computing (SBAC-PAD)}}, pages 75--82, Columbia University,
  New York, {\'E}tats-Unis, October 2012.

\bibitem{intel11:cilkplus}
{Intel Corporation}.
\newblock {Intel® Cilk™ Plus} language extension specification---version
  1.1, 2011.

\bibitem{kravets2009:graphite-opencl}
Alexey Kravets, Alexander Monakov, and Andrey Belevantsev.
\newblock {GRAPHITE-OpenCL}: Generate {OpenCL} code from parallel loops.
\newblock In {\em {Proceedings of the {GCC} Developers' Summit}}, Ottawa,
  Ontario, Canada, October 2010.

\bibitem{lementec11:xkaapi-api}
Fabien Le~Mentec, Thierry Gautier, and Vincent Danjean.
\newblock {T}he {X}-{K}aapi's {A}pplication {P}rogramming {I}nterface. {P}art
  {I}: {D}ata {F}low {P}rogramming.
\newblock Technical Report RT-0418, {Inria}, December 2011.

\bibitem{lee10:xcalablemp}
Jinpil Lee and Mitsuhisa Sato.
\newblock Implementation and performance evaluation of {XcalableMP}: A parallel
  programming language for distributed memory systems.
\newblock In {\em 39th International Conference on Parallel Processing
  Workshops ({ICPPW})}, pages 413--420, September 2010.

\bibitem{lee12:xcalablemp-gpu}
Jinpil Lee, MinhTuan Tran, Tetsuya Odajima, Taisuke Boku, and Mitsuhisa Sato.
\newblock An extension of {XcalableMP} {PGAS} lanaguage for multi-node {GPU}
  clusters.
\newblock In {\em {Euro-Par 2011}: Parallel Processing Workshops}, volume 7155
  of {\em Lecture Notes in Computer Science}, pages 429--439. Springer Berlin
  Heidelberg, 2012.

\bibitem{runtime12:starpu-handbook}
Antoine Lucas, Cyril Roeland, Cédric Augonnet, Ludovic Courtès, Nathalie
  Furmento, Olivier Aumage, and Samuel Thibault.
\newblock {StarPU} handbook, 2012.

\bibitem{amp1.0:2012}
{Microsoft Corporation.}
\newblock {C++ AMP}: Language and programming model---version 1.0, August 2012.

\bibitem{openacc1.0:2012}
{OpenACC Consortium}.
\newblock The {OpenACC™} application programming interface---version 1.0,
  November 2011.

\bibitem{pop:inria-00551518}
Antoniu Pop and Albert Cohen.
\newblock Preserving high-level semantics of parallel programming annotations
  through the compilation flow of optimizing compilers.
\newblock In {\em Proceedings of the 15th Workshop on Compilers for Parallel
  Computers ({CPC'10})}, Vienna, Austria, July 2010.

\bibitem{Rinard:1998:DIE:291889.291893}
Martin~C. Rinard and Monica~S. Lam.
\newblock The design, implementation, and evaluation of {Jade}.
\newblock {\em ACM Trans. Program. Lang. Syst.}, 20(3):483--545, May 1998.

\bibitem{Rinard:1992:HPP:147877.148003}
Martin~C. Rinard, Daniel~J. Scales, and Monica~S. Lam.
\newblock Heterogeneous parallel programming in {Jade}.
\newblock In {\em Proceedings of the 1992 ACM/IEEE conference on
  Supercomputing}, Supercomputing '92, pages 245--256, Los Alamitos, CA, USA,
  1992. IEEE Computer Society Press.

\bibitem{stallman13:gcc-manual}
Richard~M. Stallman and {the GCC Developers}.
\newblock {\em Using the {GNU Compiler Collection}}.
\newblock {GNU Press}, Boston, Massachussets, {USA}, 2012.

\bibitem{Topcuouglu:2002:PLT:566137.566142}
Haluk Topcuouglu, Salim Hariri, and Min-you Wu.
\newblock Performance-effective and low-complexity task scheduling for
  heterogeneous computing.
\newblock {\em IEEE Trans. Parallel Distrib. Syst.}, 13(3):260--274, March
  2002.

\bibitem{upc1.2-2005}
{UPC Consortium}.
\newblock {UPC} language specifications---version 1.2, May 2005.

\end{thebibliography}

\end{document}